# Magnetic topological insulator MnBi$_6$Te$_{10}$ with zero-field ferromagnetic state and gapped Dirac surface states

Shangjie Tian[1,#], Shunye Gao[2,3,#], Simin Nie[4,#], Yuting Qian[2,3], Chunsheng Gong[1], Yang Fu[1], Hang Li[2,3], Wenhui Fan[2,3], Peng Zhang[5], Takesh Kondo[5,6], Shik Shin[5,6], Johan Adell[7], Hanna Fedderwitz[7], Hong Ding[2,3,8,9], Zhijun Wang[2,3,*], Tian Qian[2,8,*], and Hechang Lei[1,*]

[1] *Department of Physics and Beijing Key Laboratory of Opto-electronic Functional Materials & Micro-nano Devices, Renmin University of China, Beijing 100872, China*

[2] *Beijing National Laboratory for Condensed Matter Physics and Institute of Physics, Chinese Academy of Sciences, Beijing 100190, China*

[3] *University of Chinese Academy of Sciences, Beijing 100049, China*

[4] *Department of Materials Science and Engineering, Stanford University, Stanford, CA, 94305, USA*

[5] *Institute for Solid State Physics, University of Tokyo, Kashiwa, Chiba 277-8581, Japan*

[6] *AIST-UTokyo Advanced Operando-Measurement Technology Open Innovation Laboratory (OPERANDO-OIL), Kashiwa, Chiba 277-8581, Japan*

[7] *MAX IV Laboratory, Lund University, P.O. Box 118, 221 00 Lund, Sweden*

[8] *Songshan Lake Materials Laboratory, Dongguan, Guangdong 523808, China*

[9] *CAS Center for Excellence in Topological Quantum Computation, University of Chinese Academy of Sciences, Beijing 100049, China*

[#] These authors contributed to this work equally.
Corresponding authors: wzj@iphy.ac.cn, tqian@iphy.ac.cn, hlei@ruc.edu.cn

**Abstract**

Magnetic topological insulators (TIs) with nontrivial topological electronic structure and broken time-reversal symmetry exhibit various exotic topological quantum phenomena. The realization of such exotic phenomena at high temperature is one of central topics in this area. We reveal that MnBi$_6$Te$_{10}$ is a magnetic TI with an antiferromagnetic ground state below 10.8 K whose nontrivial topology is manifested by Dirac-like surface states. The ferromagnetic axion insulator state with $Z_4 = 2$ emerges once spins polarized at field as low as 0.1 T, accompanied with saturated anomalous Hall resistivity up to 10 K. Such a ferromagnetic state is preserved even external field down to zero at 2 K. Theoretical calculations indicate that the few-layer ferromagnetic MnBi$_6$Te$_{10}$ is also topologically nontrivial with a non-zero Chern number. Angle-resolved photoemission spectroscopy experiments further reveal three types of Dirac surface states arising from different terminations on the cleavage surfaces, one of which has insulating behavior with an energy gap of ~ 28 meV at the Dirac point. These outstanding features suggest that MnBi$_6$Te$_{10}$ is a promising system to realize various topological quantum effects at zero field and high temperature.

Topological insulators (TIs) have attracted tremendous interests not only for demonstrating a novel classification approach of matters based on the topological invariants but also because of manifesting many of exotic phenomena protected by time-reversal symmetry, such as gapless helical states with Dirac-like dispersion on the surface or edge, and quantum spin Hall effect in two-dimensional (2D) TIs [1][2]. When introducing magnetism into TIs, the time-reversal-symmetry breaking brings even richer novel quantum effects [3][4]. A paradigm is the quantum anomalous Hall effect (QAHE), an integer QHE at zero external magnetic field, appearing in 2D Chern insulator with a non-zero Chern number. After theoretically predicted in honeycomb lattice with staggered magnetic fluxes [5], the experimental realization of QAHE has been sought for years, until it was observed in ferromagnetic (FM) TI films induced by magnetic dopants [6][7][8][9]. However, because of strong inhomogeneity of magnetic dopants, the QAHE in the magnetically doped TIs was observed at very low temperatures (usually < 100 mK), impeding the exploration of related emergent phenomena and potential technological applications.

Intrinsic magnetic TIs with periodic magnetic lattice avoid the inhomogeneity of extrinsic magnetic dopants and provide a possible route to realize high-temperature QAHE. $MnBi_2Te_4$ is the first practical example. At zero field, $MnBi_2Te_4$ is an antiferromagnetic (AFM) TI, possibly hosting exotic axion electrodynamics effects [2][10][11][12][13][14][15][16][17][18][19][20]. Importantly, down to the 2D limit, the high-field quantized Hall effect probably rooting in the topologically protected chiral edge states of Chern insulator has been predicted and demonstrated in thin $MnBi_2Te_4$ flakes [20][21][22][23][24]. However, high field is required to fully polarize the spin orientation in $MnBi_2Te_4$ in order to realize the quantization of Hall conductance. On the other hand, recent angle-resolved photoemission spectroscopy (ARPES) experiments have shown that the Dirac surface states of $MnBi_2Te_4$ remain nearly gapless in the magnetic order state [14][25][26][27]. The gapless surface states are also unfavorable to the realization of various topological quantum phenomena. Hence, it is highly desirable to obtain magnetic TIs with zero-field FM state and insulating surface states.

In this work, we demonstrate that $MnBi_6Te_{10}$ is such a magnetic TI that meet the requirements. Structurally, $MnBi_6Te_{10}$ is one of the members ($m = 1$, $n = 2$) of the homologous series of $(MnBi_2Te_4)_m(Bi_2Te_3)_n$ with stacking $m$ [Te-Bi-Te-Mn-Te-Bi-Te] septuple layers ([MnBi$_2$Te$_4$] SLs) and $n$ [Te-Bi-Te-Bi-Te] quintuple layers ([Bi$_2$Te$_3$] QLs) alternately along the $c$ axis [Fig. 1(a)] [28][29]. Although MnBi$_6$Te$_{10}$ has an AFM ground state, the interlayer AFM interaction is weakened compared to MnBi$_2$Te$_4$ due to the insertion of the QLs between the SLs [28][29][30]. Our magnetic and transport data show that the FM state emerges at field as low as 0.1 T and the saturated anomalous Hall resistivity with full spin polarization can be preserved at zero field and 2 K, which is consistent with recent study [31]. By combining theoretical calculations and ARPES experiments, we confirm that MnBi$_6$Te$_{10}$ is topologically nontrivial regardless of the magnetic states, and reveal that the topological surface states behave insulating with an energy gap of 28 meV at certain termination.

MnBi$_6$Te$_{10}$ single crystals were grown by using the self-flux method (see Supplemental Material [32]). All of peaks in the XRD pattern of a crystal can be well indexed by the (00$l$) reflections of MnBi$_6$Te$_{10}$ with the refined $c$-axis lattice parameter of 102.19(2) Å [32][33], close to the reported value of 101.985(8) Å [33]. Temperature dependence of magnetic susceptibility $\chi(T)$ at $\mu_0H$ = 5 mT for $H//c$ shows a distinct cusp at around 11 K, in contrast to the saturating behavior of $\chi(T)$ for $H//ab$. It suggests the occurrence of an AFM order at $T_N$ = 10.8 K with spin direction along the $c$ axis. At $\mu_0H$ = 1 T, the saturation of $\chi(T)$ curves for both field directions [Figs. 1(b) and 1(c)] indicates the emergence of field-induced FM state. The good fits of $\chi(T)$ curves between 20 K and 300 K at 1 T using Curie-Weiss law [32] give the effective moment $\mu_{eff}$ is 5.30(6) and 5.33(6) $\mu_B$/Mn for $H//c$ and $H//ab$, respectively, confirming the high spin state of Mn$^{2+}$ (spin-only $S$ = 5/2, $\mu_{eff}$ = 5.92 $\mu_B$) in MnBi$_6$Te$_{10}$. The fitted positive Weiss temperatures $\theta$ (= 12.3(4) and 11.4(3) K for $H//c$ and $H//ab$) indicate the dominant in-plane FM interaction in the paramagnetic (PM) state. The intralayer FM and interlayer AFM interactions results in an A-type AFM configuration below $T_N$ in MnBi$_6$Te$_{10}$. As shown in Fig. 1(d), at 2 K, the linear shape of initial magnetization $M_c(\mu_0H)$ below 0.13 T is consistent with the AFM ground state of MnBi$_6$Te$_{10}$. A spin-flip transition occurs above 0.13 T with quickly saturating at 0.2 T, i.e., entering a polarized FM state. The spin-flip-transition field $\mu_0H_{sf}$ is much lower than MnBi$_2$Te$_4$ (3.5 T - 8 T) [12][13] but comparable to MnBi$_4$Te$_7$ (0.15 T - 0.22 T) [28][29]. The smaller saturation moment $\mu_{sat}$ (~ 3.2 $\mu_B$/Mn) than $\mu_{eff}$ may be ascribed to the quenching of orbital magnetic moment induced by the crystal field in the AFM state. On the other hand, the large saturation field (~ 1.5 T) without significant hysteresis for $M_{ab}(\mu_0H)$ clearly indicates that the magnetic easy axis is along the $c$ axis. The zero-field in-plane resistivity $\rho_{xx}(T)$ exhibits a metallic behavior when $T > 20$ K [Fig. 1(e)] with a cusp centered at 11.5 K [inset of Fig. 1(e)], possibly caused by enhanced electron scattering from spin fluctuations. A sharp decrease of $\rho_{xx}(T)$ below 11.5 K reflects the weakened spin-disorder scattering due to an onset of long-range AFM magnetic order. With increasing the $c$-axial field, the cusp is suppressed and becomes a smooth drop at 1 T [32], consistent with the polarized FM state at high field.

As illustrated in Fig. 2(a), in the $M_c(\mu_0H)$ loop at $T = 2$ K, the fully polarized FM state can be persevered at zero field, remarkably different from MnBi$_2$Te$_4$ in which the FM state can only appears when $\mu_0H > 3.5$ T [10][13]. Two plateaus in the $M_c(\mu_0H)$ curve around ±0.1 T suggest an AFM state between the FM states and the non-zero moments of plateaus imply that there may be some residual FM layers in the AFM state. The linear field dependence of Hall resistivity $\rho_{xy}(\mu_0H)$ with negative slope between 0.3 T and 3 T) in the whole temperature range clearly indicates that single $n$-type carriers are dominant in MnBi$_6$Te$_{10}$ [inset of Fig. 2(b) and see more results in Ref. [32]]. The estimated carrier density $n_a$ is in the range of $4 \times 10^{20}$ - $6 \times 10^{20}$ cm$^{-3}$ [Fig. 2(g)] [32]. The anomalous Hall resistivity $\rho_{xy}^A(\mu_0H)$ obtained after subtracting the normal Hall resistivity $\rho_{xy}^N(\mu_0H)$ (= $R_0\mu_0H$) from $\rho_{xy}(\mu_0H)$ exhibits a striking similarity to $M_c(\mu_0H)$ at 2 K [Figs. 2(a) and 2(b)]. Importantly, the saturated $\rho_{xy}^A(\mu_0H)$ still can be observed when the field approaches zero, thus, unlike MnBi$_2$Te$_4$ [10][34], the zero-field AHE is truly realized in MnBi$_6$Te$_{10}$. Moreover, there is a spike-like peak in $\rho_{xy}^A(\mu_0H)$ curve at $|\mu_0H| = 0.01$ T. It could be due to the real-space topological Hall effect (THE), originating from non-coplanar spin texture with non-zero scalar spin chirality when the spins flip from a polarized FM state to an AFM state [35]. Such THE has been observed in pyrochlore Nd$_2$Mo$_2$O$_7$ and SrRuO$_3$/SrIrO$_3$ multilayer films [36][37]. As shown in Fig. 2(c), the butterfly shape of $\rho_{xx}(\mu_0H)$ is well consistent with the hysteresis of $M(\mu_0H)$ loop [Fig. 2(a)] and it changes dramatically during the spin-flip process. Moreover, the kinks at ~ 0.01 T could reflect the influence of possible spin texture on the $\rho_{xx}(\mu_0H)$.

In Fig. 2(d), while the zero-field $\rho_{xy}^A(\mu_0H)$ decreases to nearly zero quickly at higher temperatures, similar to the $M(\mu_0H)$ curves [32], the saturated values of $\rho_{xy}^A(\mu_0H = 0.1$ T) are almost unchanged even the temperature is close to $T_N$ (= 10.8 K), distinctly different from MnBi$_4$Te$_7$ where the saturated value of $\rho_{xy}^A(\mu_0H)$ diminishes rapidly when $T > 5$ K [29]. When $T > T_N$, the $\rho_{xy}^A(\mu_0H)$ decreases quickly and becomes almost zero at 14 K [Figs. 2(d) and 2(f)]. In Fig. 2(e), $\rho_{xx}(\mu_0H)$ shows a plateau in the AFM state between -0.03 T and +0.03 T, and correspondingly the $M(\mu_0H)$ [32] and $\rho_{xy}^A(\mu_0H)$ become very low. On the other hand, in the polarized FM state, the $\rho_{xx}(\mu_0H)$ shows a negative magnetoresistance (MR), partially ascribed to the suppression of spin-disorder scattering. The negative MR becomes positive one at higher fields or temperatures [32], possibly due to the dominance of normal positive orbital MR.

In order to illuminate the topological properties of MnBi$_6$Te$_{10}$, we carried out first-principles calculations [32]. An "open core" treatment of Mn 3$d$ electrons is used to treat them as core states. The band structure with spin-orbit coupling (SOC) in the PM state shows a full band gap at the Fermi level ($E_F$) [Fig. 3(a)]. Based on the Fu-Kane parity criterion, the computed $Z_2$ indices are 1;(000), which indicate that it is a strong TI in the high-temperature PM state. In contrast, the AFM state emerges at low temperature, and there are two Mn atoms with different spin orientations in one AFM unit cell. In the AFM configuration, the spatial inversion symmetry $P$ (with the origin located at a Mn atom) is preserved. Even though the time-reversal symmetry $\Theta$ is broken, the two kinds of Mn atoms can be related by a half translation of the lattice vector (i.e. $T_{1/2} = [\vec{a} + \vec{b} + \vec{c}]/2$) in the $z$ direction, where $\vec{a}$, $\vec{b}$ and $\vec{c}$ are primitive lattice vectors, respectively. In other words, the anti-unitary symmetry $S = \Theta T_{1/2}$ is respected. In the band structure of the AFM state, all the bands are doubly degenerate due to the presence of the combined anti-unitary symmetry with the condition $(PS)^2 = -1$ for every $k$-point. In the $k_z = 0$ plane, there is an AFM $Z_2$ invariant ($v_{AFM0}$) due to $S^2 = -1$. In the presence of inversion symmetry $P$, the AFM $Z_2$ can be further simplified as follows:

$$(-1)^{v_{AFM0}} \equiv \prod_{i=1}^{4} \prod_{n}^{n_{occ}/2} \xi_{2n}(K_i)$$

Explicitly, $K_1 = (0, 0, 0)$; $K_2 = (0.5, -0.5, 0)$; $K_3 = (0.5, 0, -0.5)$; $K_4 = (0, 0.5, -0.5)$, where ($u$, $v$, $w$) are given in units of primitive reciprocal lattice vectors. $K_2$, $K_3$, $K_4$ are related by $C_{3z}$. The parity eigenvalues for occupied Kramers pairs of bands are given in Table I. The obtained $v_{AFM0} = 1$ suggests that it is an AFM TI with a full bulk gap in Fig. 3(b), resulting in gapless surface states on the side surfaces where the anti-unitary symmetry $S$ is preserved.

Most interestingly, the band structure in the FM state also shows a gap [~ 0.15 eV, Fig. 3(c)]. In the FM state preserving inversion symmetry, the parity-based invariant is defined by [38][39]:

$$Z_4 = \sum_{\alpha=1}^{8} \sum_{n=1}^{n_{occ}} \frac{1 + \xi_n(\Lambda_\alpha)}{2} \, mod \, 4$$

Based on the computed parity eigenvalues of occupied states at eight inversion-symmetry-invariant momenta (only four of them are distinct) in Table I, the obtained $Z_4 = 2$ indicates that the FM MnBi$_6$Te$_{10}$ is an axion insulator with $\theta = \pi$. Here, the coefficient $\theta$ is defined in the field theory description of the topological magnetoelectric (TME) effect $S_\theta = \frac{\theta e^2}{4\pi^2} \int dt d^3 x \, \boldsymbol{E} \cdot \boldsymbol{B}$ with $\boldsymbol{E}$ and $\boldsymbol{B}$ electromagnetic fields [4][17]. The strain tolerance of the nontrivial topology in the FM MnBi$_6$Te$_{10}$ has also been checked. The nontrivial topology is robust with respect to both the hydrostatic expansion and the hydrostatic compression [32]. Moreover, we have performed the calculations for the FM slab structures of different layers. The results show that two, three and four trilayers (one trilayer stands for the sandwich structure of [Bi$_2$Te$_3$]-[MnBi$_2$Te$_4$]-[Bi$_2$Te$_3$]) are intrinsic Chern insulators. The electronic structure of the four-trilayer slab is presented in Fig. 3(d), with an energy gap of 20 meV. Its Chern number is obtained to be -1 from the Wilsonloop calculations in the inset of Fig. 3(d) (see more results in Ref. [32]). As a result, the QAHE can be expected in few-layer MnBi$_6$Te$_{10}$ with no external field.

To confirm the topological properties of MnBi$_6$Te$_{10}$, we use ARPES to measure the electronic structures on the (001) surface [32]. The synchrotron ARPES data in Fig. 4(a) show that the MnBi$_6$Te$_{10}$ samples are electron doped, consistent with the negative Hall coefficient. The conduction bands lie above -0.2 eV and the valence bands lie below -0.4 eV, forming a band gap of around 0.2 eV, in agreement with the band calculation. The most remarkable feature in Fig. 4(a) is the existence of Dirac surface states within the band gap, which is compelling evidence of the nontrivial topology of MnBi$_6$Te$_{10}$. We then carry out systematic ARPES experiments using the 7-eV laser as incident light. We scan the electronic structures at different locations on the cleavage surface by shifting the sample, and identify three types of surface states, labelled as I, II, and III, respectively [Figs. 4(b)-4(e)]. Considering the van der Waals heterostructure of MnBi$_6$Te$_{10}$, cleavage may occur between two Bi$_2$Te$_3$ layers or between MnBi$_2$Te$_4$ and Bi$_2$Te$_3$ layers, resulting in three kinds of terminations, namely *BBA*, *BAB* and *ABB* (*A* and *B* represent the [MnBi$_2$Te$_4$] and [Bi$_2$Te$_3$] layers, respectively), as shown in Fig. 4(f). The three types of surface states should derive from different terminations. The type-II and III surface states appear to have a gapless Dirac point [Figs. 4(d) and 4(e)], whereas the type-I surface states open an energy gap of ~ 28 meV at the Dirac point [Fig. 4(c)]. The surface states are almost unchanged when undergoing the AFM transition (See Supplementary Material [32]), similar to the observation in MnBi$_2$Te$_4$ and MnBi$_4$Te$_7$ [25][26][27].

Since the combined symmetry *S* is not preserved at the (001) surface in the AFM state, the (001) Dirac surface states are expected to open a finite energy gap. To understand the dependence of the gap size on the terminations, we compute the surface states at different terminations by constructing two AFM slabs with a thickness of thirty layers with a vacuum of ~ 15 Å. As shown in Fig. 4(f), the first slab has the same termination *BAB* on both sides, while the second one has the termination *BBA* on one side and *ABB* on the other. The calculated results in Fig. 4(g) show that the energy gap of the surface states at $\bar{\Gamma}$ is 100 meV, 29.7 meV and 0.2 meV at the terminations *ABB*, *BAB*, and *BBA*, respectively. The farther the [MnBi$_2$Te$_4$] layer are from the termination, the smaller the band gap of the surface Dirac cone is. The variation of the gap size is associated with the magnitude of the surface exchange field [28], which is derived from the Mn 3*d* magnetic moments. The effect of the surface exchange field on the surface states is significantly suppressed when the [MnBi$_2$Te$_4$] layer is away from the termination. We infer that the type-I surface states with an observable gap probably arise from the termination *ABB*. We note that the calculated surface states are not well consistent with the ARPES results, such as the gap size and band dispersions. Further theoretical and experimental studies, such as STM measurements, are needed to definitely determine the correspondence between the surface states and terminations. Although the origin of the gap is still unclear at the current stage [40][41], the gapped surface state at a particular kind of termination is favorable to realize the relevant topological quantum phenomena, such as QAHE.

In summary, MnBi$_6$Te$_{10}$ exhibits a nontrivial topology regardless of the magnetic states. The A-type AFM ground state can be tuned to the FM axion insulator state under field as low as 0.1 T and the latter one can be preserved without any degradation at 2 K at zero field and up to 10 K at 0.1 T. $MnBi_6Te_{10}$ exhibits termination-dependent surface states, one of which opens a large gap of ~ 28 meV. Moreover, theoretical calculations show that the Chern insulator state is expected in the 2D limit. Because of the easily accessible low-field and high-temperature FM state as well as the insulating surface states, $MnBi_6Te_{10}$ provides a very promising platform to realize the QAHE as well as other exotic topological quantum effects at high temperature and nearly zero field.

**Acknowledgments**

This work was supported by the National Key R&D Program of China (Grants No. 2018YFE0202600, 2016YFA0300504, 2016YFA0300600, 2016YFA0401000), the National Natural Science Foundation of China (No. 11574394, 11504117, 11774423, 11822412, 11622435, U1832202, 11888101), the Fundamental Research Funds for the Central Universities, and the Research Funds of Renmin University of China (RUC) (15XNLQ07, 18XNLG14, 19XNLG17, 20XHN062), the Chinese Academy of Sciences (QYZDB-SSW-SLH043 and XDB28000000), the Beijing Municipal Science & Technology Commission (Z171100002017018 and Z181100004218005). Laser ARPES work was supported by JSPS KAKENHI (Grants No. JP18H01165 and JP19H00651) and Grant-in-Aid for JSPS Fellows (Grant No. 19F19030). Z.J.W. acknowledges support from the National Thousand-Young-Talents Program, the CAS Pioneer Hundred Talents Program, and the National Natural Science Foundation of China (No. 11974395).

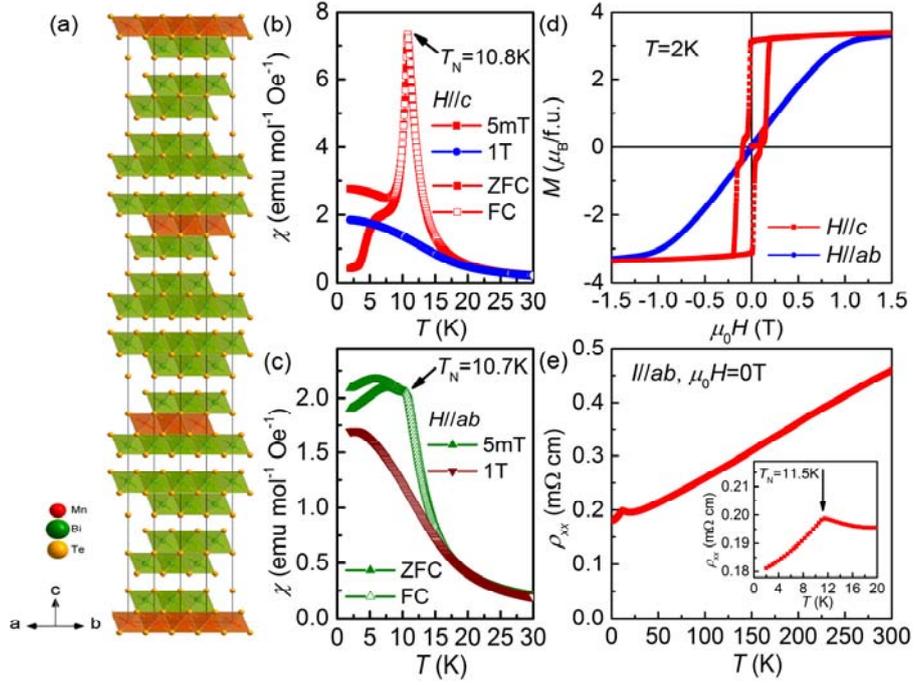

Figure 1. (a) Crystal structure of MnBi$_6$Te$_{10}$ with space group *R-3m*. (b, c) Temperature dependence of $\chi(T)$ at $\mu_0H$ = 5 mT and 1 T along the *c* axis and *ab* plane, respectively. (d) $M(\mu_0H)$ curves for *H//c* and *H//ab* at *T* = 2 K. (e) Temperature dependence of $\rho_{xx}(T)$ at zero field. Inset: enlarged part of $\rho_{xx}(T)$ at low-temperature region.

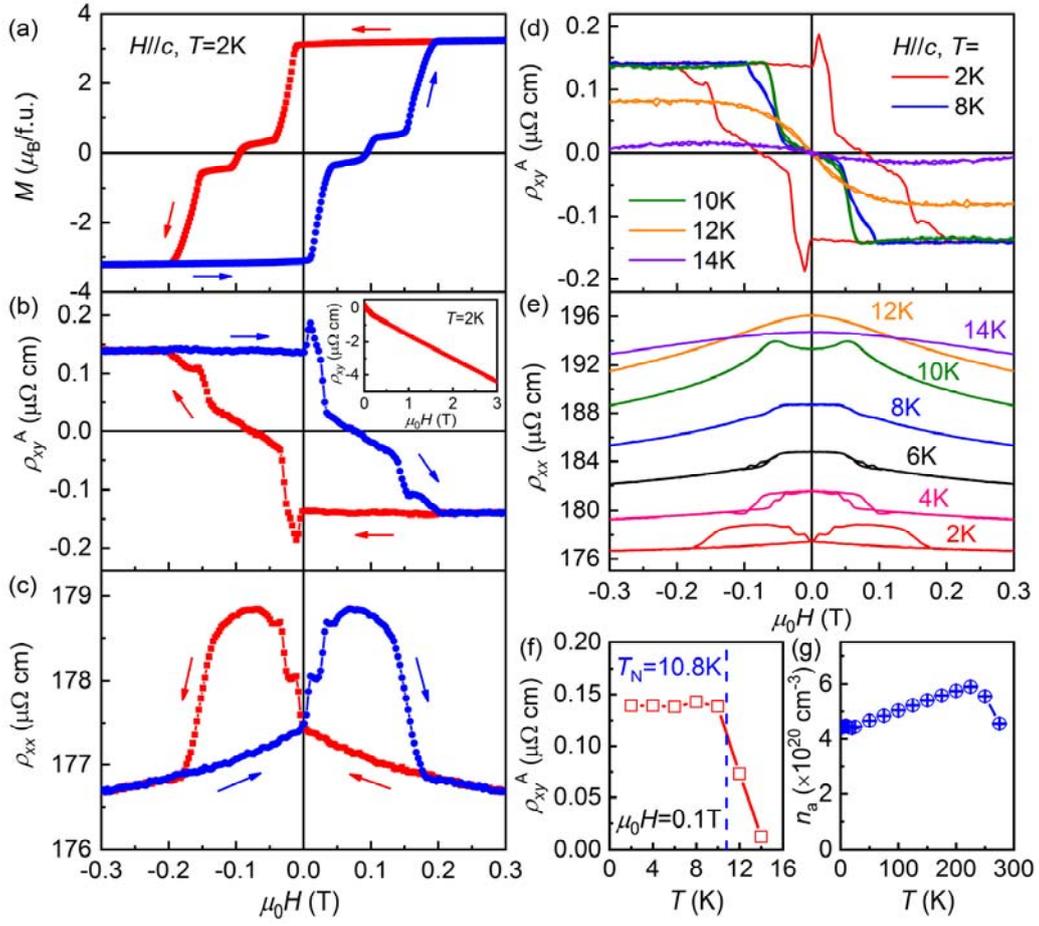

Figure 2. Field dependence of (a) $M(\mu_0 H)$, (b) $\rho_{xy}^A(\mu_0 H)$ and (c) $\rho_{xx}(\mu_0 H)$ at $T$ = 2 K and $\mu_0 H$ up to ±0.3 T. (d) $\rho_{xy}^A(\mu_0 H)$ and (e) $\rho_{xx}(\mu_0 H)$ at various temperatures with -0.3 T ≤ $\mu_0 H$ ≤ 0.3 T. (f) Temperature dependence of $\rho_{xy}^A(\mu_0 H = 0.1$ T). The blue dashed line denotes the $T_N$. (g) Temperature dependence of $n_a$.

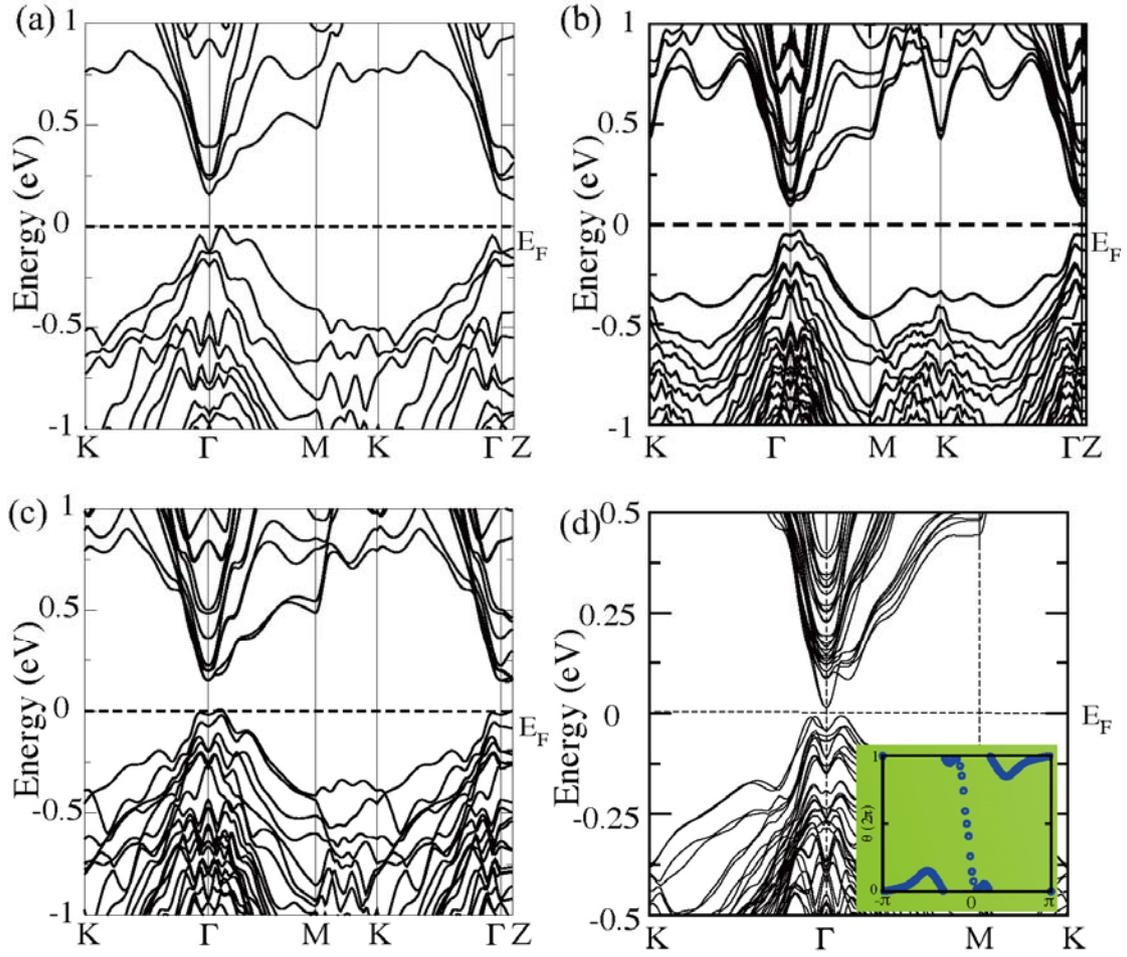

Figure 3. Band structures for the (a) PM, (b) AFM, and (c) FM states of MnBi$_6$Te$_{10}$ along high symmetry lines. The high-symmetry *k*-points are labeled in the corresponding conventional Brillouin zones (BZs). (d) Electronic structure of the FM slab of four trilayers with an energy gap of 20 meV. Inset: the corresponding Wilsonloop calculations with a nontrivial Chern number (*C* = -1).

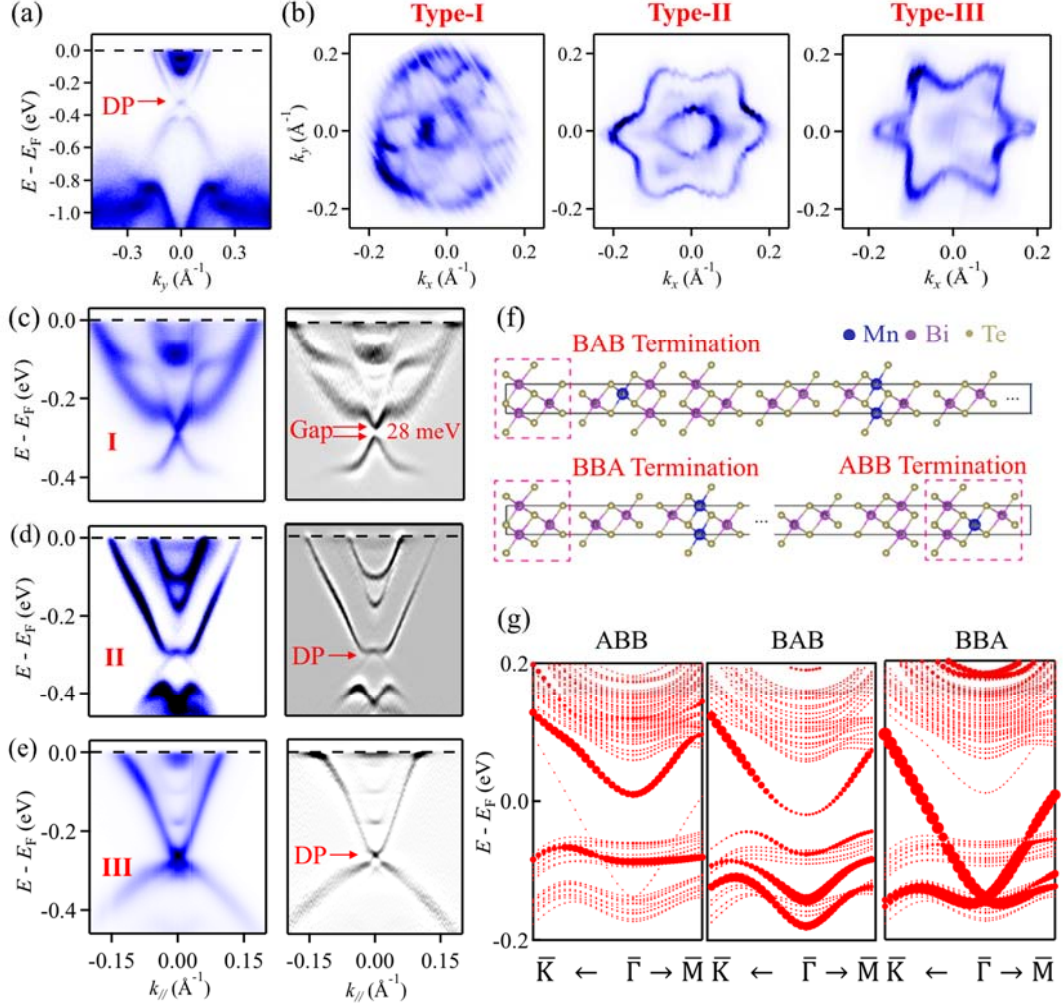

Figure 4. Dirac surface states on the (001) surface of MnBi$_6$Te$_{10}$. (a) ARPES intensity plot along the cut across $\bar{\Gamma}$ measured at $hv$ = 21 eV and 47 K. (b) Fermi surface intensity maps of three types of electronic structures taken with the 7-eV laser at 40 K. (c, d, e) ARPES intensity plots (left) and their second derivative with respect to energy (right) along the cut across $\bar{\Gamma}$ of three types of electronic structures. (f) Schematic plots are side views of two AFM slabs with three kinds of terminations. (g) Calculated band structures around $\bar{\Gamma}$ with projected weight of the topmost layer (dashed boxes in (f)) on terminations *ABB* (left), *BAB* (middle), and *BBA* (right), respectively. The weight is represented by the size of the red circles.

| $K_1$ | (0, 0, 0) | (0.5, -0.5, 0) | (0.5, 0, -0.5) | (0, 0.5, -0.5) |
|---|---|---|---|---|
| AFM | 36, 29 | 35, 30 | 35, 30 | 35, 30 |
| $\Lambda_\alpha$ | (0, 0, 0) | (0.5, 0, 0) | (0.5, 0.5, 0) | (0.5, 0.5, 0.5) |
| PM | 16, 14 | 15, 15 | 15, 15 | 15, 15 |
| FM | 37, 28 | 35, 30 | 35, 30 | 35, 30 |

Table I. Numbers of occupied bands of odd and even parity at the corresponding TRIM points. (*u, v, w*) are the coordinates of *k*-point with respect to the primitive reciprocal vectors. "*p, q*" indicate the numbers of (the numbers of pairs of) even-parity and odd-parity occupied bands for the FM state (for the PM and AFM states), respectively. The total numbers of occupied bands for the PM, FM and AFM states are 60, 75, and 130, respectively.